\documentclass[11pt]{article}

\usepackage{amsmath,amssymb,amsthm}
\usepackage{graphicx}
\usepackage{hyperref}
\usepackage{natbib}
\usepackage{xcolor}
\usepackage{booktabs}
\usepackage{float}
\usepackage{geometry}
\usepackage{authblk}
\usepackage{lineno}  
\usepackage{mathtools}

\title{Risk-Adjusted Performance of Random Forest Models in High-Frequency Trading}

\author[1]{Akash Deep\thanks{Email: akash.deep@ttu.edu}}
\author[2]{Abootaleb Shirvani}
\author[1]{Chris Monico}
\author[1]{Svetlozar Rachev}
\author[3]{Frank J. Fabozzi}

\affil[1]{Department of Mathematics and Statistics, Texas Tech University, USA}
\affil[2]{Department of Mathematical Sciences, Kean University, USA}
\affil[3]{Johns Hopkins Carey Business School, Johns Hopkins University, USA}

\date{\today}

\renewenvironment{abstract}{
  \centerline{\large\bf Abstract}
  \begin{quote}
}{\end{quote}}

\begin{document}

\maketitle

\begin{abstract}
Because of the theoretical challenges posed by the Efficient Market Hypothesis to technical analysis, the effectiveness of technical indicators in high-frequency trading remains inadequately explored, particularly at the minute-level frequency, where effects of the microstructure of the market dominate.
This study evaluates the integration of traditional technical indicators with random forest regression models using minute-level SPY data, analyzing 13 distinct model configurations.
Our empirical results reveal a stark contrast between in-sample and out-of-sample performance, with $R^2$ values deteriorating from 0.749--0.812 during training to negative values in testing.
A feature importance analysis demonstrates that primary price-based features dominate the predictions made by the model, accounting for over 60\% of the importance, while established technical indicators, such as RSI and Bollinger Bands, account for  only 14\%--15\%.
Although the indicator-enhanced models achieved superior risk-adjusted metrics, with Rachev ratios between 0.919 and 0.961, they consistently underperformed a simple buy-and-hold strategy, generating returns ranging from -2.4\% to -3.9\%.
These findings challenge conventional assumptions about the usefulness of technical indicators in algorithmic trading, suggesting that in high-frequency contexts, they may be more relevant to risk management rather than to predicting returns.
For practitioners and researchers, our findings indicate that successful high-frequency trading strategies should focus on adaptive feature selection and regime-specific modeling rather than relying on traditional technical indicators, as well as indicating the critical importance of robust out-of-sample testing in the development of a model.
\end{abstract}


\section{Introduction}

Accurate prediction of the stock market remains a fundamental yet highly challenging objective in financial research due to the volatility, noise, and stochasticity in financial markets.
As \citet{aldridge2013high} point out, the increasing prevalence of high-frequency trading (HFT), where trades are executed within milliseconds, has intensified the demand for predictive models that can rapidly adapt to market fluctuations and structural complexity.
However, developing such models requires overcoming significant hurdles, including the inherent noise in high-frequency data and the rapid shifts in market sentiment, as documented by \citet{gu2020empirical}.

A central theoretical debate in financial economics revolves around the effectiveness of technical indicators in predicting prices.
The Efficient Market Hypothesis (EMH) proposed by \citet{fama1970efficient} suggests that asset prices fully incorporate all available information, rendering historical price-based signals ineffective for forecasting.
However, the continued widespread use of technical indicators by traders raises questions about the validity of this assumption, particularly in short-term, high-frequency contexts.
As \citet{barberis2003survey} point out, technical analysis may capture behavioral biases, such as herding and overconfidence, that contribute to transient market inefficiencies being manifested in the prices.
These biases are particularly pronounced at the minute-level, where noise traders (market participants who rely on historical patterns rather than fundamental analysis) may introduce temporary mispricings that machine learning models could exploit.

Machine learning (ML) has emerged as a powerful tool for predicting stock prices, enabling the identification of nonlinear dependencies and complex relationships within historical data.
Traditional statistical methods, such as autoregressive integrated moving average (ARIMA) and generalized autoregressive conditional heteroskedasticity (GARCH) models, often struggle to capture the intricate price dynamics of high-frequency markets, due to their assumptions of linearity.
By contrast, ML models, such as random forest regression (RFR), support vector regression (SVR), and gradient boosting, have demonstrated improved predictive performance in financial applications \citep{derbentsev2020machine}.
However, their effectiveness is highly contingent on the feature selection, particularly in high-frequency trading environments where the dominance of market noise presents a formidable challenge.

Technical analysis, as outlined by \citet{murphy1999technical}, employs historical price and volume data through indicators such as Bollinger Bands, exponential moving averages (EMA), and the Commodity Channel Index (CCI), to detect trends and signal potential price reversals.
These indicators aim to reflect aggregate market sentiment and trader behavior.
However, their effectiveness in HFT is still  debated.
Studies such as \citet{abrol2016high} suggest that traditional technical indicators often generate unreliable signals in high-frequency environments, where rapid price fluctuations introduce significant noise.
Although recent research has examined the integration of technical indicators with machine learning models \citep{zanc2019forecasting, fischer2018deep}, much of this work has focused on daily or hourly data, leaving the complexities of minute-level stock price movements relatively unexplored \citep{zhang2010high}.

In the present paper, we assess the predictive and risk management performance of random forest regression models augmented with technical indicators for high-frequency stock price prediction.
Building on previous research that primarily focuses on daily or hourly data, we extend the analysis to minute-level data, incorporating advanced risk-adjusted performance metrics.
This allows us to examine the interplay between technical indicators and the effects of the microstructure of the market, providing new insights into their role in high-frequency trading.

This study tests the hypothesis that while incorporating technical indicators can improve risk-adjusted performance, their effectiveness at prediction diminishes in volatile, high-frequency environments where noise dominates the signal.
In particular, we expect that primary price-based features will contribute more significantly to the predictions of the model than the technical indicators will, aligning with prior evidence for their limited predictive power in short-term trading.
In addition, we evaluate the alignment of our findings with the Efficient Market Hypothesis (EMH) by analyzing in-sample versus out-of-sample performance.
Our findings provide empirical support for the semi-strong form of the EMH, that while technical indicators may be able to briefly exploit market inefficiencies, their predictive power is limited.

Unlike many of the existing studies, which primarily evaluate technical indicators in daily or hourly trading, our paper is among the first to systematically assess their effectiveness at the minute level, a granularity where the effects of microstructures of the market and noise dominate.
Furthermore, prior studies have predominantly relied on conventional evaluation metrics, such as root mean squared error (RMSE) and R-squared ($R^2$), which provide limited insight into risk-adjusted performance.
In contrast, our study employs advanced risk--reward measures, including the Rachev ratio and the gains--loss ratio, offering a more comprehensive evaluation in high-frequency contexts of trading strategies based on machine learning \citep{cheridito2013reward}.
By combining insights from technical analysis, machine learning, behavioral finance, and advanced risk management, this paper provides actionable implications for both academics and practitioners seeking to refine predictive modeling techniques for financial markets.

Our findings indicate that while technical indicator-augmented models obtain superior risk-adjusted metrics, when it comes to generating excess returns, they perform worse than a simple buy-and-hold strategy.
Following the framework established by \citet{barberis2003survey}, our results suggest that while technical indicators can enhance risk management, they may not provide sufficient predictive power to consistently outperform baseline strategies in high-frequency environments dominated by market noise and sentiment-driven trading behavior.
These insights emphasize the importance of selective feature engineering, regime-aware modeling, and adaptive risk management techniques in the application of machine learning to financial markets, so as to improve the stability of the predictions in high-frequency contexts.

\section{Literature Review}

Predicting stock prices has been a long-standing challenge due to the volatility and complexity of the market.
The Efficient Market Hypothesis (EMH) suggests that prices fully reflect all available information, leaving little room for prediction \citep{fama1970efficient}.
However, behavioral finance research has identified systematic deviations from market efficiency, particularly in high-frequency contexts where noise traders may rely heavily on technical indicators \citep{barberis2003survey}.
This tension between rational finance and behavioral finance provides a motivation for evaluating the effectiveness of such technical indicators, as noise traders lacking access to fundamental data may disproportionately rely on technical indicators, potentially creating temporary market inefficiencies \citep{shleifer1997limits}.

Advances in machine learning (ML) and the increasing availability of high-frequency trading (HFT) data have made possible the empirical investigation of these theoretical predictions.
Traditional econometric models, such as ARIMA and GARCH, were initially used for forecasting stock prices, but often struggled with nonstationary data and volatility clustering, as noted by \citet{zhang1998forecasting} and \citet{patel2015predicting}.
These limitations pointed to the need for integrating ML techniques with traditional financial models in order to improve their predictive accuracy.

In the mid-1990s, ensemble methods, such as random forest, were developed, demonstrating robustness in handling high-dimensional datasets and reducing overfitting through bagging \citep{ho1995random}.
Recent studies have demonstrated the role of ML in enabling adaptive strategic behaviors on the part of high-frequency traders.
By leveraging tools like genetic algorithms, traders can process complex information about the microstructure of the market and optimize their trading strategies in real time, significantly enhancing their profitability under varying conditions \citep{arifovic2022machine}.
The interaction between the speed of the trading and the efficiency of the market has also been explored, finding a hump-shaped relation between speed, efficiency, and the profitability of the trader.

By the early 2000s, studies like \citet{bollinger2002bollinger} began exploring technical indicators, such as Bollinger Bands (BBs), to gauge market trends and overbought or oversold conditions.
Meanwhile, the Commodity Channel Index (CCI) and Exponential Moving Average (EMA) emerged as widely used tools for capturing short-term price movements \citep{lambert1983commodity, murphy1999technical}.
However, the standalone use of these indicators often yielded inconsistent results, particularly in noisy and volatile environments, such as HFT \citep{zhang2010high}.

As ML techniques advanced, studies in the 2010s began integrating technical indicators with ML models to improve the predictive performance.
For instance, \citet{fischer2018deep} demonstrated that combining technical indicators with LSTM networks could reduce noise in high-frequency stock data and enhance the accuracy of the predictions.
\citet{gu2020empirical} expanded on this by showing that ML can uncover market inefficiencies, though these tend to be temporary and limited in nature.
Their work emphasized the importance of robust out-of-sample testing and careful feature selection in predictive modeling.

Despite these advances, challenges, such as overfitting and generalization, remained.
Researchers like \citet{agrawal2019stock} emphasized the importance of domain-specific feature selection to mitigate these problems, while \citet{lim2021time} emphasized the need for dynamic models capable of adapting to changing market conditions.
\citet{akyildirim2023forecasting} demonstrated that Random Forest models excel at identifying nonlinear patterns in data and perform consistently across different time scales, making them particularly suitable for high-frequency stock price forecasting, where complex relationships exist.

By the early 2020s, the focus shifted toward hybrid strategies combining multiple technical indicators and ML techniques.
\citet{zanc2019forecasting} explored the integration of BBs with LSTM networks, showing improvements in predictive accuracy under volatile market conditions.
At the same time, studies began addressing the limitations of traditional evaluation metrics, such as the Sharpe and Sortino ratios, which often assume that the returns are normally distributed.
Advanced risk--reward metrics, such as the Rachev and modified Rachev ratios, were introduced to provide a more nuanced understanding of how a model performs in volatile environments \citep{cheridito2013reward}.

Recent work has increasingly focused on high-frequency data and its unique challenges.
\citet{kearns2013machine} highlighted the difficulties of extracting meaningful signals from noisy HFT data, while \citet{o2015high} emphasized that market microstructure takes on heightened importance at very fast speeds.
These studies underscored the need for models that balance predictive power with robustness against market noise.

Despite substantial progress, significant gaps remain in the literature.
Much of the existing work has focused on lower-frequency data, leaving minute-level and tick-level observations underexplored \citep{zhang2010high}.
Advanced risk--reward metrics, though proposed, have seen limited application in HFT contexts.
Hybrid strategies combining multiple technical indicators have shown promise, but their incremental benefits over simpler models are not well-documented.
Generalization challenges persist, particularly in HFT settings, where fleeting arbitrage opportunities and high levels of noise increase the risk of overfitting.

This paper contributes to the field by systematically evaluating the predictive and risk-adjusted performance, in an HFT context, of random forest regression models combined with technical indicators. 
It incorporates advanced risk--reward metrics to provide a comprehensive assessment of model performance.
This paper also addresses generalization issues through rigorous validation techniques and highlights the limited utility of technical indicators in highly volatile settings.
By combining technical analysis, ML, and risk management, this paper offers actionable insights for practitioners and researchers aiming to refine predictive modeling in financial markets.

\section{Method}

In this section, we describe the data acquisition process, the computation of the technical indicators, the ML model (random forest regressor), and the trading simulation framework.
The decisions made at each step are guided by the need to rigorously assess the impact of technical indicators on stock price prediction using random forest.

\subsection{Data Acquisition and Preprocessing}

The dataset used in this study consists of minute-level historical stock data for the SPY (S\&P 500 ETF), covering the period from April 2024 to September 2024.
The data includes essential fields such as the opening, high, low, and close prices, as well as the trading volume, for each minute.
The data were obtained from the Bloomberg Terminal, ensuring high accuracy and reliability \citep{bloomberg_data}.
Each data point is timestamped in Central Time (CT), and the dataset covers the typical US stock market hours from 9:30 AM to 4:00 PM Eastern Time (ET), adjusted for daylight savings time.

Additionally, the 10-year US Treasury yield is incorporated as a proxy for the risk-free rate, a crucial factor in calculating excess returns.
These data are reported daily and were also sourced from the Bloomberg Terminal, spanning the same time frame as the SPY data \citep{pastor2003liquidity}.

\subsubsection{Log Returns and Volatility}
To normalize the stock price data and reduce the effects of scale, we compute the log returns for the opening, high, low, and close prices.
Log returns are preferred in financial time series due to their ability to capture percentage changes and handle volatility over time \citep{box2015time}.
The log return for a price series \( P_t \) is
\begin{linenomath}
\begin{equation}
\text{log\_return}_t = \log\left(\frac{P_t}{P_{t-1}}\right),
\end{equation}
\end{linenomath}
where \( P_t \) is the price at time \( t \).
This process is also defined for  the opening, high, low, and closing prices, with the resulting log returns stored as additional columns in the dataset.
Additionally, we compute rolling $Z$-scores for the trading volume to capture anomalies in the volume.
The rolling $Z$-score of the volume is \citep{box2015time}
\begin{linenomath}
\begin{equation}
\text{volz}_t = \frac{\text{volume}_t - \text{mean}(\text{volume})}{\text{std}(\text{volume})},
\end{equation}
\end{linenomath}
where the mean and standard deviation are computed over a rolling window of 60 minutes.

The 10-year US Treasury yield, provided on a daily basis, is used to compute a per-minute risk-free rate, which is necessary for calculating excess returns.
For any minute $t$ within a trading day $d$, the transformation from the daily yield to a per-minute rate is given by
\begin{linenomath}
\begin{equation}
r_{\text{per-minute}}(t) = \left(1 + r_{\text{daily}}(d)\right)^{\frac{1}{1440}} - 1,
\end{equation}
\end{linenomath}
where \( 1440 \) is the number of minutes in a day and \(r_{\text{daily}}(d)\) is the daily risk-free rate derived from the most recently available Treasury yield prior to day $d$.
This ensures that each minute's risk-free rate reflects the prevailing daily rate for its trading day.

\subsubsection{Data Filtering and Splitting}
The dataset is filtered to focus on regular market trading hours, between 10:00 AM and 3:30 PM CT, to avoid periods of low liquidity, such as pre-market and after-hours trading \citep{mcgroarty2019high}.
The filtered dataset is then split into training and testing sets, with 80\% of the data allocated to training and 20\% to testing.
The splitting is time-ordered to preserve the temporal nature of the stock price data and avoid data leakage.

The processed dataset is used for computing a set of technical indicators, which serve as input features for the ML models described in subsequent sections.
The computed technical indicators include the simple moving average (SMA), EMA, moving average convergence divergence (MACD), Relative Strength Index (RSI), and others, as detailed below.

\subsection{System Architecture}
\begin{figure}[H]
\centering
\includegraphics[width=\textwidth]{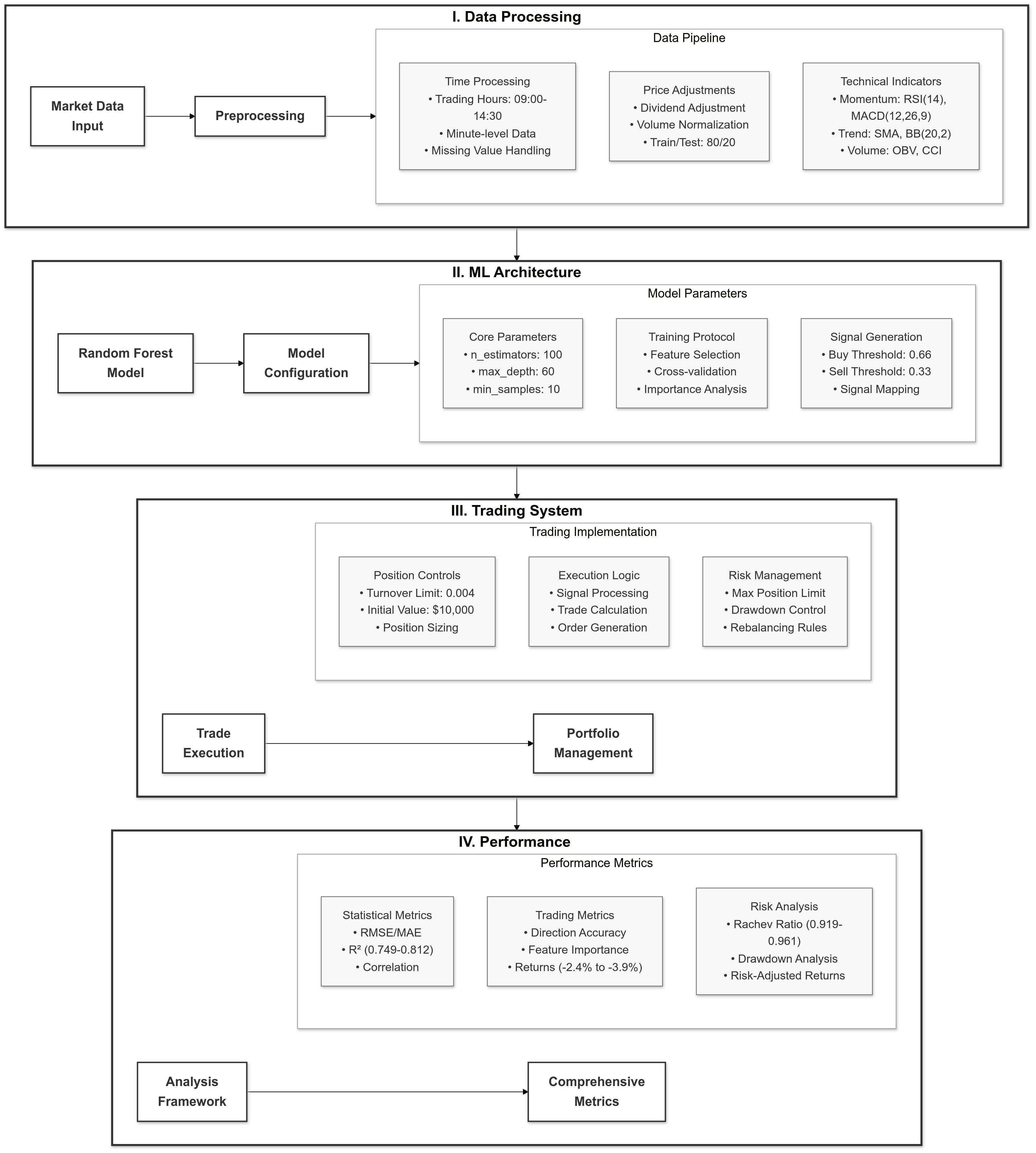}
\caption{Architecture of the ML-based trading system.
There are four integrated phases: (I) Data Processing: handling minute-level SPY data (09:00--14:30), incorporating dividend adjustments, and computing technical indicators including RSI(14), MACD(12,26,9), and Bollinger Bands(20,2); (II) ML Architecture: Random Forest implementation with the specific hyperparameters (n\_estimators=100, max\_depth=60) and quantile-based signal generation (buy: 0.66, sell: 0.33); (III) Trading System: real-time position management with turnover constraints (0.004) and initial capital allocation (\$10,000); and (IV) Performance Analysis: comprehensive evaluation using statistical metrics (the range of values of $R^2$ is 0.749--0.812) and risk-adjusted measures (Rachev ratio: 0.919--0.961).
This system is an example of a practical integration of traditional technical analysis with modern ML approaches, while emphasizing risk management and computational efficiency in high-frequency trading contexts.}
\label{fig:system_architecture}
\end{figure}

Figure \ref{fig:system_architecture} presents a comprehensive view of our ML-based trading system, illustrating the interconnections between the processing of the data, the development of the model, and the evaluation of the performance.

Our implementation follows a systematic approach where the data preprocessing feeds into the ML pipeline, which in turn feeds into the trading decisions.
This framework incorporates comprehensive risk management and performance evaluation, ensuring robust validation of the strategy's effectiveness.
Each component is optimized for high-frequency trading, with particular attention to computational efficiency and real-time processing.

\subsection{Technical Indicators}
To capture diverse aspects of market behavior, we selected a set of widely recognized technical indicators, each chosen for its unique contribution to predicting price movements or managing risk.
These technical indicators encompass a variety of trend-following, momentum, and volume-based metrics, enabling a robust, multi-faceted analysis of minute-level price movements.

For instance, the EMA and MACD offer insights into the strength and direction of a trend, whereas BBs and the RSI gauge the volatility and overbought/oversold conditions, respectively \citep{murphy1999technical}.
The average directional index (ADX) measures the robustness of a trend, the on-balance volume (OBV) measures the volume flow, and the CCI detects cyclical price movements \citep{lambert1983commodity}.
By combining these technical indicators, we aimed to create a feature set capable of reflecting both short-term and long-term market dynamics, thus enhancing the predictive accuracy and enabling nuanced risk management \citep{zanc2019forecasting}.

Unless otherwise specified, all non-trivial technical indicator formulae presented are derived from the seminal work \citep{murphy1999technical}.

\subsubsection{Simple Moving Average (SMA)}
The SMA smooths price data by averaging the closing prices over a window of \(N\) periods:
\begin{linenomath}
\begin{equation}
\text{SMA}_{N,t} = \frac{1}{N} \sum_{i=0}^{N-1} C_{t-i},
\end{equation}
\end{linenomath}
where \(C_t\) denotes the closing price at time \(t\).
In our implementation, the current price is normalized by the SMA:
\begin{linenomath}
\begin{equation}
\hat{\text{SMA}}_{N,t} = \frac{C_t}{\text{SMA}_{N,t}}.
\end{equation}
\end{linenomath}
This ensures scale invariance and helps the model better learn from the price data.

\subsubsection{Exponential Moving Average (EMA)}
The EMA places more weight on recent prices, making it more responsive to changes in prices.
It is calculated recursively as follows:
\begin{linenomath}
\begin{equation}
\text{EMA}_t = \alpha C_t + (1 - \alpha) \text{EMA}_{t-1},
\end{equation}
\end{linenomath}
where \( \alpha = \frac{2}{N+1} \) is the smoothing factor for a window size \(N\).
In our implementation, the EMA is normalized similarly to the SMA:
\begin{linenomath}
\begin{equation}
\hat{\text{EMA}}_t = \frac{C_t}{\text{EMA}_t}.
\end{equation}
\end{linenomath}
This ratio stabilizes the feature and makes it more useful for prediction.

\subsubsection{Moving Average Convergence Divergence (MACD)}
The MACD measures the difference between short-term and long-term EMAs.
It is computed as follows:
\begin{linenomath}
\begin{equation}
\text{MACD}_t = \text{EMA}_{12,t} - \text{EMA}_{26,t}.
\end{equation}
\end{linenomath}
The signal line \( \text{SIG}_t \) is a 9-period EMA of the MACD line.
We use the following ratio to normalize the MACD:
\begin{linenomath}
\begin{equation}
r_{\text{MACD},t} = \frac{\text{MACD}_t - \text{SIG}_t}{0.5 \left( |\text{MACD}_t| + |\text{SIG}_t| \right)}.
\end{equation}
\end{linenomath}
This ensures that large fluctuations in the MACD do not overwhelm the model.

\subsubsection{Relative Strength Index (RSI)}
The RSI is a momentum oscillator that measures the speed and change of price movements \citep{wilder1978new}.
It is computed as follows:
\begin{linenomath}
\begin{equation}
\text{RSI}_t = 100 - \frac{100}{1 + \frac{\text{avg\_gain}_t}{\text{avg\_loss}_t}},
\end{equation}
\end{linenomath}
where \( \text{avg\_gain}_t \) and \( \text{avg\_loss}_t \) are the exponentially smoothed averages of the gains and losses over a window of 14 periods.
The RSI ranges from 0 to 100, identifying possible overbought and oversold conditions.

\subsubsection{Bollinger Bands (BBs)}
Bollinger bands are volatility bands placed two standard deviations above and below a moving average.
They are defined by
\begin{linenomath}
\begin{equation}
\text{UBB}_t = \text{SMA}_{N,t} + 2\sigma_t, \quad \text{LBB}_t = \text{SMA}_{N,t} - 2\sigma_t,
\end{equation}
\end{linenomath}
where \( \sigma_t \) is the standard deviation of the prices over the last \(N\) periods \citep{bollinger2002bollinger}.
The normalized BB percentage is
\begin{linenomath}
\begin{equation}
\text{BB\%}_t = \frac{C_t - \text{LBB}_t}{\text{UBB}_t - \text{LBB}_t}.
\end{equation}
\end{linenomath}
This captures where the price sits within the volatility bands.

\subsubsection{Stochastic Oscillator (SO)}
The stochastic oscillator (SO) measures the relative position of the closing price compared to the high--low range over a specified period (typically 14 periods).
It is computed as follows:
\begin{linenomath}
\begin{equation}
\%K_t = 100 \times \frac{C_t - L_{14,t}}{H_{14,t} - L_{14,t}},
\end{equation}
\end{linenomath}
where \( L_{14,t} \) and \( H_{14,t} \) denote the lowest and highest prices over the last 14 periods.
The slow stochastic oscillator \( \%D_t \) is a 3-period moving average of \( \%K_t \).

\subsubsection{Fibonacci Retracement (Fib)}
The Fibonacci retracement is used to identify potential support and resistance levels in a price trend.
For a window \(N\), the retracement level is
\begin{linenomath}
\begin{equation}
R_t = \frac{H_{N,t} - C_t}{H_{N,t} - L_{N,t}},
\end{equation}
\end{linenomath}
where \(H_{N,t}\) and \(L_{N,t}\) are the highest and lowest prices over the window.
We use common Fibonacci levels (0.236, 0.382, 0.500, 0.618, 0.764) to identify potential reversal points.

\subsubsection{Average Directional Index (ADX)}
The ADX measures the strength of a trend, regardless of its direction.
The ADX is derived from the directional movement indicators \(DI^+_t\) and \(DI^-_t\):
\begin{linenomath}
\begin{equation}
\text{ADX}_t = \frac{|DI^+_t - DI^-_t|}{DI^+_t + DI^-_t}.
\end{equation}
\end{linenomath}
The directional movement indicators \( DI^+_t \) and \( DI^-_t \) are normalized by the average true range (ATR).

\subsubsection{On-Balance Volume (OBV)}
The OBV is a cumulative indicator that sums the volumes, depending on whether the price is rising or falling:
\begin{linenomath}
\begin{equation}
\text{OBV}_t = \text{OBV}_{t-1} + \text{sgn}(C_t - C_{t-1}) V_t,
\end{equation}
\end{linenomath}
where \(V_t\) is the trading volume at time \(t\), and the signum function determines the direction of the volume flow.

\subsubsection{Windowed Relative OBV (WROBV)}
The windowed relative OBV (WROBV) is a modified version of the OBV. It is the weighted sum of the values, for a rolling window of size \(N\), of the cumulative OBV.  This smooths out the indicator:
\begin{linenomath}
\begin{equation}
\text{WROBV}_t = \frac{\sum_{i=0}^{N-1} \text{OBV}_{t-i}}{\sum_{i=0}^{N-1} V_{t-i}}.
\end{equation}
\end{linenomath}
This rolling normalization prevents the OBV from becoming excessively large, and focuses on recent price--volume dynamics.

\subsubsection{Commodity Channel Index (CCI)}
The CCI measures the deviation of the typical price from its moving average:
\begin{linenomath}
\begin{equation}
p_t = \frac{H_t + L_t + C_t}{3}.
\end{equation}
\end{linenomath}
The CCI is given by
\begin{linenomath}
\begin{equation}
\text{CCI}_t = \frac{p_t - \text{SMA}_{N,t}}{0.015 \times \text{MAD}_t},
\end{equation}
\end{linenomath}
where \( \text{MAD}_t \) is the mean, over a rolling window of size \(N\),  of the absolute deviations of \(p_t\).

\subsubsection{Ichimoku Cloud (Ichimoku)}
The Ichimoku Cloud is a comprehensive technical indicator that provides a holistic view of support, resistance, the direction of the trend, and momentum \citep{patel2010trading}.
It has five main components:
\begin{itemize}
\item Tenkan-sen (Conversion Line): This line is a short-term indicator calculated as the midpoint of the highest high and the lowest low over the past \( N \) periods:
\begin{linenomath}
\begin{equation}
\text{Tenkan}_t = \frac{\max(H_{t-N}, \ldots, H_t) + \min(L_{t-N}, \ldots, L_t)}{2},
\end{equation}
\end{linenomath}
where \(H_t\) and \(L_t\) denote the high and low prices at time \(t\), respectively.
Typically, \( N = 9 \).

\item Kijun-sen (Base Line): The base line is a longer-term indicator calculated similarly to the Tenkan-sen but over a longer window \( M \):
\begin{linenomath}
\begin{equation}
\text{Kijun}_t = \frac{\max(H_{t-M}, \ldots, H_t) + \min(L_{t-M}, \ldots, L_t)}{2}.
\end{equation}
\end{linenomath}
This line provides a measure of medium-term momentum, with \( M = 26 \) being a common value.

\item Senkou Span A (Leading Span A): Senkou Span A is the midpoint between the Tenkan-sen and Kijun-sen, plotted \( M \) periods ahead:
\begin{linenomath}
\begin{equation}
\text{Senkou A}_t = \frac{\text{Tenkan}_t + \text{Kijun}_t}{2} \quad \text{(shifted forward by \( M \) periods)}.
\end{equation}
\end{linenomath}
This span, along with the following Senkou Span B, forms the Ichimoku Cloud.

\item Senkou Span B (Leading Span B): This span is the midpoint of the highest high and lowest low over the past \( L \) periods and is also plotted \( M \) periods ahead:
\begin{linenomath}
\begin{equation}
\text{Senkou B}_t = \frac{\max(H_{t-L}, \ldots, H_t) + \min(L_{t-L}, \ldots, L_t)}{2} \quad \text{(shifted forward by \( M \) periods)}.
\end{equation}
\end{linenomath}
The area between Senkou Span A and Senkou Span B is shaded to form the `cloud,' which can act as dynamic support or resistance.

\item Chikou Span (Lagging Span): The Chikou Span is the current closing price plotted \( M \) periods in the past:
\begin{linenomath}
\begin{equation}
\text{Chikou}_t = C_t \quad \text{(shifted backward by \( M \) periods)}.
\end{equation}
\end{linenomath}
This line provides a lagging indication of price action and helps confirm the direction of a trend.
\end{itemize}
The Ichimoku Cloud provides a visual representation of support and resistance, the direction of a trend, and momentum.
The interaction between the price and the cloud helps identify potential reversals or continuations in the trend.
In our implementation, we calculate all five components of the Ichimoku cloud and incorporate the leading spans (Senkou A and Senkou B) as features in the machine learning model.

\subsection{Random Forest and Validation}

The underlying predictive model in our framework uses a random forest regressor (RFR), which is an ensemble learning method that aggregates predictions from multiple decision trees to capture complex, non-linear relations in high-frequency financial data \citep{ho1995random, breiman2001random, buitinck2013api}.
Let $\mathcal{D} = \{(\mathbf{x}_i, y_i)\}_{i=1}^n$ denote our training dataset, where $\mathbf{x}_i \in \mathbb{R}^p$ denotes the feature vector consisting of technical indicators and price-based features at time $i$, and $y_i \in \mathbb{R}$ denotes the corresponding log return.

The RFR constructs an ensemble of $B$ decision trees, where each tree $T_b$ is trained on a bootstrap sample $\mathcal{D}_b$ drawn with replacement from $\mathcal{D}$.
For a given input vector $\mathbf{x}$, the model's prediction is
\begin{linenomath}
\begin{equation}
\hat{f}(\mathbf{x}) = \frac{1}{B} \sum_{b=1}^{B} T_b(\mathbf{x}),
\end{equation}
\end{linenomath}
where $T_b(\mathbf{x})$ denotes the prediction of the $b$-th tree.
Each individual tree is constructed by recursively partitioning the feature space to minimize the mean squared error (MSE):
\begin{linenomath}
\begin{equation}
\text{MSE}(t) = \frac{1}{|\mathcal{D}_t|} \sum_{i \in \mathcal{D}_t} (y_i - \bar{y}_t)^2,
\end{equation}
\end{linenomath}
where $\mathcal{D}_t$ denotes the set of training samples at node $t$, and $\bar{y}_t$ is the mean response value in node $t$.

Our implementation employs scikit-learn's RandomForestRegressor with the following parametrization:
\begin{linenomath}
\begin{equation}
\Theta = \{\theta_B, \theta_d, \theta_s, \theta_f, \theta_l, \theta_r\},
\end{equation}
\end{linenomath}
where $\theta_B = 100$ (n\_estimators), $\theta_d = 60$ (max\_depth), $\theta_s = 10$ (min\_samples\_split), $\theta_f = \text{'log2'}$ (max\_features), $\theta_l = 1$ (min\_samples\_leaf), and $\theta_r = 42$ (random\_state).
This configuration performs cross-validation through Out-of-Bag (OOB) sampling \citep{hastie2009random}, where approximately one-third of the observations are automatically held out during the training of each tree, serving as a built-in validation set.

To determine the signal, we employ a quantile-based thresholding mechanism.
Let $\hat{f}(\mathbf{x}_t)$ be the model's prediction at time $t$.
During training, we compute threshold values $q_{0.33}$ and $q_{0.66}$, which are the 33rd and 66th percentiles of the model's predictions on the training set.
The signal function $s: \mathbb{R} \rightarrow \{\text{"sell"}, \text{"hold"}, \text{"buy"}\}$ is defined by
\begin{linenomath}
\begin{equation}
s(\hat{f}(\mathbf{x}_t)) = \begin{cases}
\text{"buy"} & \text{if } \hat{f}(\mathbf{x}_t) \geq q_{0.66} \\
\text{"hold"} & \text{if } q_{0.33} < \hat{f}(\mathbf{x}_t) < q_{0.66} \\
\text{"sell"} & \text{if } \hat{f}(\mathbf{x}_t) \leq q_{0.33}
\end{cases}
\end{equation}
\end{linenomath}

The importance of a feature is computed using the mean decrease in impurity across all trees:
\begin{linenomath}
\begin{equation}
I_j = \frac{1}{B} \sum_{b=1}^B \sum_{t \in \mathcal{T}_b} \Delta \text{MSE}_{t,j} \mathbb{1}(v(t)=j),
\end{equation}
\end{linenomath}
where $\mathcal{T}_b$ is the set of nodes in tree $b$, $v(t)$ is the feature used for splitting at node $t$, and $\Delta \text{MSE}_{t,j}$ is the decrease in MSE achieved by splitting on feature $j$ at node $t$.

For temporal validation, we employ a chronological partitioning:
\begin{linenomath}
\begin{equation}
\mathcal{D}_{\text{train}} = \{(\mathbf{x}_i, y_i)\}_{i=1}^{\lfloor 0.8n \rfloor}, \quad
\mathcal{D}_{\text{test}} = \{(\mathbf{x}_i, y_i)\}_{i=\lfloor 0.8n \rfloor + 1}^n,
\end{equation}
\end{linenomath}
ensuring strict temporal ordering and preventing look-ahead bias.
This 80--20 split, combined with the OOB error estimation, provides a robust validation framework that appropriately handles both the ensemble nature of Random Forests and the sequential characteristics of high-frequency trading data.
As demonstrated by \citet{hastie2009random}, the OOB error estimate is nearly equivalent to leave-one-out cross-validation, providing an unbiased estimate of the test error and making additional $k$-fold cross-validation unnecessary.

While time-series cross-validation (TSCV), such as rolling or walk-forward validation, is a common approach in the forecasting of financial time-series, its application to high-frequency trading (HFT) remains computationally intensive.
Given the ensemble nature of Random Forests and the strict temporal partitioning employed in our study, we rely on Out-of-Bag (OOB) error estimation as an efficient alternative.
This method maintains the chronological integrity of the data while avoiding excessive computational overhead.
Future research should explore the trade-off between computational feasibility and the robustness benefits of TSCV, particularly in adaptive trading models where market regimes shift dynamically.

\subsection{Trading Simulation}

We simulate a trading strategy based on the buy, sell, and hold signals generated by the random forest model.
The trading simulation starts with an initial value of \$10,000.
The following actions are taken based on the predictions of the model:

\begin{itemize}
    \item \textbf{Buy Signal}: If the model predicts an upward price movement, a portion of the available cash is used to buy shares.
    \item \textbf{Sell Signal}: If a downward price movement is predicted, a portion of the holdings is sold.
    \item \textbf{Hold Signal}: If no significant price movement is predicted, no action is taken.
\end{itemize}

To approximate real-world trading constraints, we impose a turnover constraint limiting position changes in any minute to 0.4\% of the portfolio value:
\begin{linenomath}
\begin{equation}
\frac{\text{value\_traded}}{\text{portfolio\_value}} \leq 0.004.
\end{equation}
\end{linenomath}

This constraint was chosen to align with SPY's typical daily turnover rate of approximately 3\%.
By limiting per-minute turnover to 0.4\%, our simulation ensures that total daily changes  in the position remain within realistic bounds, given the ETF's observed liquidity characteristics.
All trades are executed at minute-end closing prices.
While this implementation provides realistic control of the sizes of the positions, it is somewhat optimistic, as it does not take into account bid--ask spreads, commission costs, or potential price impact.
These limitations should be considered when interpreting the performance of the strategy.

\subsection{Evaluation Metrics}

We assess the performance of the random forest model using a comprehensive set of metrics that evaluate both the accuracy  of the predictions  and the risk-adjusted returns:

\begin{itemize}
    \item \textbf{Root Mean Squared Error (RMSE)}: this is the square root of the average of the squares of the differences between the predicted and actual returns:
    \begin{linenomath}
    \begin{equation}
    \text{RMSE} = \sqrt{\frac{1}{n} \sum_{i=1}^{n} (\hat{y}_i - y_i)^2}.
    \end{equation}
    \end{linenomath}
    A lower RMSE indicates a better accuracy.

    \item \textbf{Mean Absolute Error (MAE)}: This is the average of the absolute differences between the predicted and actual returns, offering an intuitive measure of the model's accuracy.

    \item \textbf{R-squared ($R^2$)}: This is the proportion of the variance in the target variable explained by the model:
    \begin{linenomath}
    \begin{equation}
    R^2 = 1 - \frac{\sum_{i=1}^{n} (y_i - \hat{y}_i)^2}{\sum_{i=1}^{n} (y_i - \bar{y})^2},
    \end{equation}
    \end{linenomath}
    where \( \bar{y} \) is the mean of the actual returns.
A higher \( R^2 \) indicates a better performance of the  model.

    \item \textbf{Trend Accuracy}: This evaluates the model's ability to predict the direction (up or down) of price movements:
    \begin{linenomath}
    \begin{equation}
    \text{Trend Accuracy} = \frac{1}{n} \sum_{i=1}^{n} \mathbf{1}(\text{sign}(y_i) = \text{sign}(\hat{y}_i)),
    \end{equation}
    \end{linenomath}
    where \( \mathbf{1} \) is an indicator function returning 1 if the predicted direction matches the actual direction, and 0 otherwise.

    \item \textbf{Sharpe Ratio}: This assesses the risk-adjusted performance of the trading strategy:
    \begin{linenomath}
    \begin{equation}
    \text{Sharpe Ratio} = \frac{\mathbb{E}[r_p - r_f]}{\sigma_p},
    \end{equation}
    \end{linenomath}
    where \( r_p \) is the asset return, \( r_f \) is the risk-free rate, and \( \sigma_p \) is the standard deviation of the returns.
A higher Sharpe ratio indicates better risk-adjusted returns.

    \item \textbf{Maximum Drawdown}: This is the largest peak-to-trough decline in the value of the asset over the testing period:
    \begin{linenomath}
    \begin{equation}
    \text{Max Drawdown} = \max_{t \in T} \left(\frac{\text{peak}_t - \text{trough}_t}{\text{peak}_t}\right).
    \end{equation}
    \end{linenomath}
    This metric is crucial for evaluating the worst-case performance of the trading strategy during periods of market stress.

    \item \textbf{Sortino Ratio}: This is a variant of the Sharpe ratio that focuses only on downside risk:
    \begin{linenomath}
    \begin{equation}
    \text{Sortino Ratio} = \frac{\mathbb{E}[r_p - r_f]}{\text{Downside Deviation}},
    \end{equation}
    \end{linenomath}
    where the downside deviation is calculated using only negative returns.
This ratio penalizes excessive downside risk more than overall volatility.
\end{itemize}

These metrics provide a well-rounded evaluation of the RFR's predictive accuracy and its ability to manage risk in the context of an HFT strategy.
\subsection{Selection of Risk--Reward Ratios}

In selecting risk--reward ratios for this study, we follow the theoretical framework laid out by \citet{cheridito2013reward}, focusing on ratios that satisfy the following four critical properties:

\begin{itemize}
    \item \textbf{Monotonicity (M):} This property ensures that the reward--risk ratio (RRR) increases as returns increase, for a fixed level of risk.
Essentially, this criterion reflects the intuitive idea that `more is better.' Formally, for two random variables \( X \) and \( Y \), where \( X \geq Y \), we should have \( \rho(X) \geq \rho(Y) \).
    \item \textbf{Quasi-Concavity (Q):} Quasi-concavity encourages diversification, ensuring that the ratio prefers averages over extremes.
If a reward--risk ratio satisfies this property, this means that a diversified portfolio will generally be preferred over a concentrated risk.
Formally, for random variables \( X \) and \( Y \), and for any \( \lambda \in [0, 1] \), we should have \( \rho(\lambda X + (1 - \lambda) Y) \geq \min(\rho(X), \rho(Y)) \).
    \item \textbf{Scale Invariance (S):} Scale invariance means that the ratio remains unchanged when both the return and the risk of a portfolio are scaled by the same factor.
This ensures that the ratio is consistent across different investment sizes; it requires that \( \rho(\lambda X) = \rho(X) \) for all positive scalars \( \lambda \).
    \item \textbf{Distribution-based (D):} This property ensures that the ratio depends only on the distribution of the returns \( X \) and not on any specific realization of \( X \).
This is essential for generalizing the performance metric across different scenarios and portfolio strategies.
\end{itemize}

These properties form a robust basis for evaluating performance metrics, ensuring that they promote diversification and reward consistency.
Many risk--reward ratios used in the financial literature—such as the Sharpe ratio, Sortino ratio, Rachev ratio, and others—naturally satisfy these criteria.
The ratios chosen for this study align with these principles, allowing a comprehensive evaluation of the performance of a portfolio.

\begin{table}[H]
\centering
\caption{Risk--reward ratios used in the study.}
\footnotesize 
\renewcommand{\arraystretch}{1.5} 
\setlength{\tabcolsep}{7pt} 
\begin{tabular}{lcp{7cm}}
\toprule
\textbf{Ratio} & \textbf{Formula} & \textbf{Description} \\ \hline
Sharpe ratio & 
$\large \frac{\mathbb{E}[R_p - R_f]}{\sigma_p}$ & 
$R_p$: Portfolio return, $R_f$: Risk-free rate, $\sigma_p$: Standard deviation of excess returns.\newline
Measures the excess return per unit of risk (volatility), highlighting risk-adjusted performance.\\ \hline

Sortino ratio & 
$\large \frac{\mathbb{E}[R_p - R_f]}{\sigma_d}$ & 
$\sigma_d$: Standard deviation of negative returns (downside risk).\newline
Improves on the Sharpe ratio by focusing only on downside risk, penalizing large losses more than fluctuations from gains.\\ \hline

Rachev ratio & 
$\large \frac{\mathbb{E}[R_p \mid R_p \geq \text{VaR}_{1-\gamma}]}{\mathbb{E}[R_p \mid R_p \leq \text{VaR}_{\beta}]}$ & 
$VaR$: Value-at-Risk, $\gamma$: Upper quantile, $\beta$: Lower quantile.\newline
Measures tail risk by comparing the potential gains in the best-case scenario with the worst-case losses.\\ \hline

Modified Rachev ratio & 
$\large \frac{\mathbb{E}[R_p \mid R_p \geq \text{VaR}_{1-\delta}] / \epsilon}{\mathbb{E}[R_p \mid R_p \leq \text{VaR}_{\delta}] / \gamma}$ & 
$\delta, \epsilon$: Additional parameters to refine the evaluation of risk.\newline
Extends the Rachev ratio to offer a more granular comparison between upper and lower tails at multiple confidence levels.\\ \hline

Distortion RRR & 
$\large \frac{\mathbb{E}[R_p \mid R_p \geq \text{VaR}_{1-\beta}]}{\mathbb{E}[R_p \mid R_p \leq \text{VaR}_{\beta}]}$ & 
$VaR$: Value-at-Risk, $\beta$: Confidence level.\newline
Uses a distortion function to adjust the weights of the gains and losses, allowing flexible risk assessments depending on the investor's preferences.\\ \hline

Gains--Loss ratio & 
$\large \frac{\mathbb{E}[R_p \mid R_p > 0]}{\mathbb{E}[|R_p| \mid R_p < 0]}$ & 
The ratio of the average positive returns over the average negative returns, providing a simple risk--reward comparison.\\ \hline

STAR ratio & 
$\large \frac{\mathbb{E}[R_p - R_f]}{\mathbb{E}[R_p \mid R_p \leq \text{VaR}_{\alpha}]}$ & 
$VaR$: Value-at-Risk, $\alpha$: Confidence level.\newline
Focuses on tail risk, using the Conditional Value-at-Risk (CVaR), also known as the expected shortfall, to take into account extreme losses.\\ \hline

MiniMax ratio & 
$\large \frac{\mathbb{E}[R_p]}{\text{Max Drawdown}}$ & 
Max Drawdown: Largest peak-to-trough decline in portfolio value.\newline
Compares the average return to the largest drawdown, focusing on how the strategy performs relative to its worst loss.\\ \hline

Gini ratio & 
$\large \frac{\sum_{i=1}^N (2i - N - 1) R_i}{N \sum_{i=1}^N R_i}$ & 
$R_i$: Sorted returns, $N$: Number of observations.\newline
Measures the inequality in the distribution of returns, analogous to the Gini coefficient used in economics.\\
\bottomrule
\end{tabular}
\end{table}

\section{Results}\label{sec:results}

This section summarizes the performance of the random forest regression (RFR) models with and without technical indicators, compares them to a buy-and-hold benchmark, and discusses their statistical significance.
The results include the predictive accuracy, the trading outcomes, the risk-adjusted performance, the contributions made by each feature, and residual analyses.

\subsection{Predictive Performance}\label{sec:predictive_performance}

\subsubsection{Training vs.\ Testing Metrics}
We trained and tested 13 RFR models, differing in their inclusion of technical indicators, and compared their performance to a buy-and-hold benchmark.
Table~\ref{tab:model_metrics} presents the root mean square error (RMSE), mean absolute error (MAE), and \(R^2\) for both training and testing sets.
Although the models generally achieved strong results in-sample (training \(R^2\) from 0.749 to 0.812), out-of-sample performance deteriorated (testing \(R^2\) in the range \(-0.020\) to \(-0.016\)).
This discrepancy points to overfitting, consistent with the challenges often encountered when applying ML to minute-level data.

\begin{table}[h]
\centering
\caption{Model performance metrics for training and testing.}
\label{tab:model_metrics}
\begin{tabular}{lcccccc}
\toprule
\textbf{Model} & \multicolumn{2}{c}{\textbf{RMSE}} & \multicolumn{2}{c}{\textbf{MAE}} & \multicolumn{2}{c}{\textbf{R\textsuperscript{2}}} \\
& \textbf{Train} & \textbf{Test} & \textbf{Train} & \textbf{Test} & \textbf{Train} & \textbf{Test} \\
\midrule
RFR (no indicators)  & 0.00021 & 0.00036 & 0.00015 & 0.00024 & 0.786 & -0.020 \\
rfr\_boll            & 0.00021 & 0.00036 & 0.00015 & 0.00024 & 0.812 & -0.016 \\
rfr\_ema             & 0.00022 & 0.00036 & 0.00016 & 0.00024 & 0.749 & -0.019 \\
rfr\_rsi             & 0.00021 & 0.00036 & 0.00015 & 0.00024 & 0.802 & -0.017 \\
\bottomrule
\end{tabular}
\end{table}

All models have comparable RMSEs (0.00036) and MAEs (0.00024) in the test set, indicating little variation in forecasting error.
The negative out-of-sample \(R^2\) values for each model confirm that high in-sample fits did not translate into predictive power on unseen data.

\subsection{Outcomes of the Trading Strategies}\label{sec:trading_outcomes}

\subsubsection{Portfolio Value and Returns}

\begin{figure}[htbp]
\centering
\includegraphics[width=\textwidth]{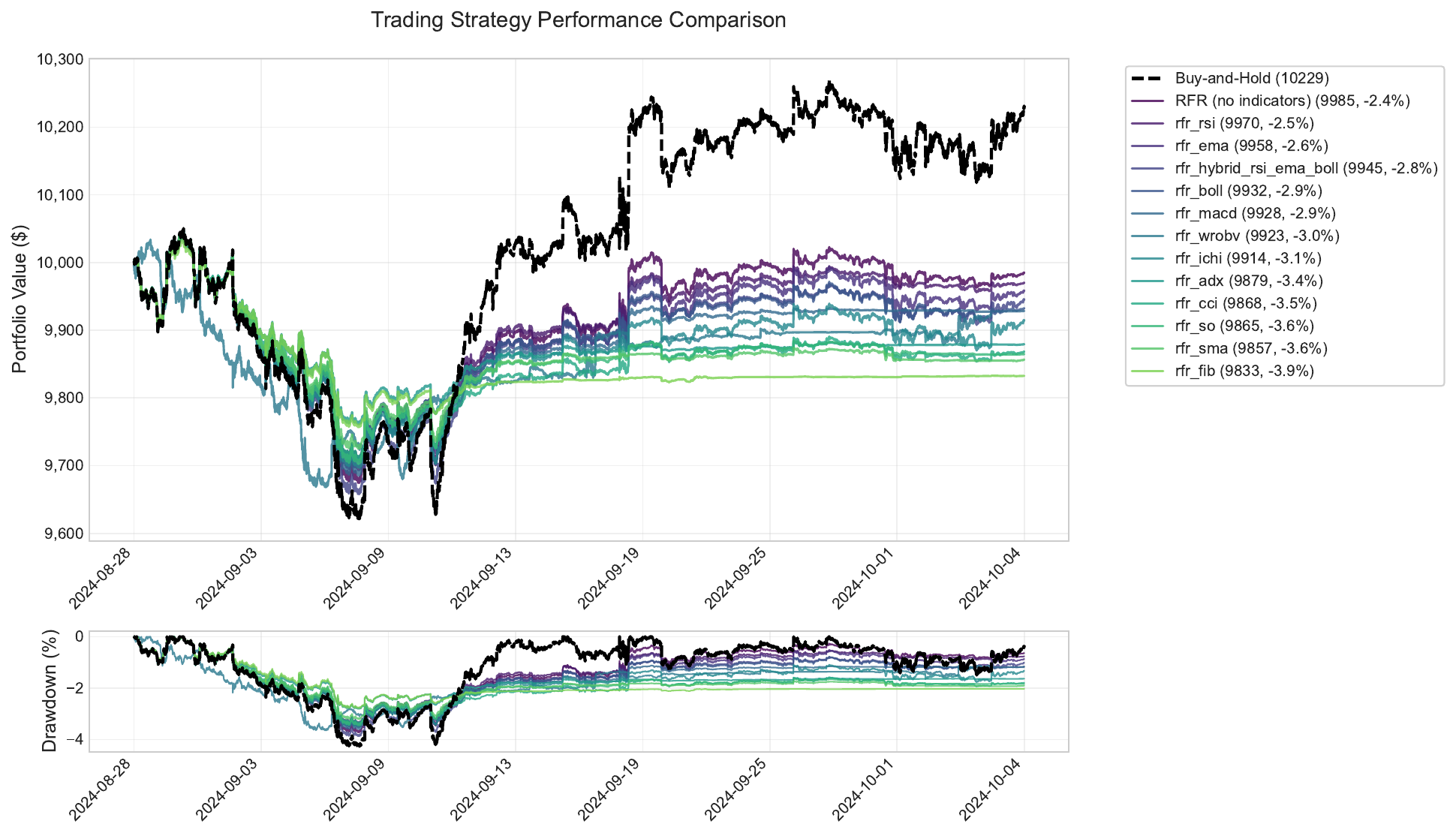}
\caption{Trajectories of the values of the portfolios for different trading strategies.
The buy-and-hold approach ended at \$10,229, while the algorithmic models underperformed to different degrees.
Maximum drawdown was around 4\%.}
\label{fig:portfolio_performance}
\end{figure}

We simulated a trading strategy for each model from August~28, 2024, to October~4, 2024, starting with \$10{,}000.
Figure~\ref{fig:portfolio_performance} shows the trajectories of the portfolios, and Table~\ref{tab:performance_summary} shows the final portfolio values, returns, and major performance ratios.
The buy-and-hold strategy ended at \$10{,}229, which counts as  a 0\% deviation from the baseline, since it is the baseline in our setting. All RFR-based strategies underperformed.

\begin{table}[H]
\centering
\caption{Summary of trading performance.}
\label{tab:performance_summary}
\begin{tabular}{lccccc}
\toprule
\textbf{Model} & \textbf{Final Value (\$)} & \textbf{Return (\%)} & \textbf{Sharpe} & \textbf{Sortino} & \textbf{Rachev} \\
\midrule
Buy-and-hold              & 10,229 & 0.00   & --       & --       & --    \\
RFR (no indicators)       & 9,985  & -2.40  & 0.0046   & 0.0047   & 0.946 \\
rfr\_rsi                  & 9,970  & -2.50  & -0.0015  & -0.0018  & 0.961 \\
rfr\_ema                  & 9,958  & -2.60  & -0.0020  & -0.0024  & 0.961 \\
rfr\_hybrid\_rsi\_ema\_boll & 9,945 & -2.80 & -0.0024 & -0.0029 & 0.956 \\
rfr\_boll                 & 9,932  & -2.90  & -0.0033  & -0.0040  & 0.957 \\
rfr\_macd                 & 9,928  & -2.90  & -0.0035  & -0.0041  & 0.953 \\
rfr\_wrobv                & 9,923  & -3.00  & -0.0041  & -0.0046  & 0.938 \\
rfr\_ichi                 & 9,914  & -3.10  & -0.0040  & -0.0048  & 0.950 \\
rfr\_adx                  & 9,879  & -3.40  & -0.0078  & -0.0089  & 0.937 \\
rfr\_cci                  & 9,868  & -3.50  & -0.0069  & -0.0082  & 0.943 \\
rfr\_so                   & 9,865  & -3.60  & -0.0073  & -0.0083  & 0.939 \\
rfr\_sma                  & 9,857  & -3.60  & -0.0082  & -0.0093  & 0.937 \\
rfr\_fib                  & 9,833  & -3.90  & -0.0116  & -0.0133  & 0.919 \\
\bottomrule
\end{tabular}
\end{table}

Although each model ended below \$10,000, a few (notably, RFR with no indicators or rfr\_rsi) performed slightly better than the others in risk-adjusted terms, with Sharpe ratios near 0.00 to 0.0046.
None, however, surpassed the buy-and-hold benchmark in absolute returns.

That RFR-based strategies behave worse than the buy-and-hold benchmark can be attributed to transaction costs and market noise, which diminish the effectiveness of short-term trading strategies.
While technical indicators provide some value in capturing short-term inefficiencies, the minute-level predictive horizon may not be sufficient to extract profitable trading signals.
Additionally, high turnover rates in algorithmic strategies increase trading costs, further eroding potential returns in real-world implementations.

Our results align with prior studies, such as \citet{peng2021feature}, which found that technical indicators provide limited predictive value in deep learning models trained on daily-level data.
Unlike our Random Forest approach, studies leveraging LSTMs and attention-based models have demonstrated better sequence-learning capabilities.
However, our findings suggest that even with alternative architectures, the predictive power for high-frequency trading intervals remains constrained due to the market's microstructural noise.

\subsection{Risk-Adjusted Performance}\label{sec:risk_adjusted_performance}

\begin{figure}[htbp]
\centering
\includegraphics[width=\textwidth]{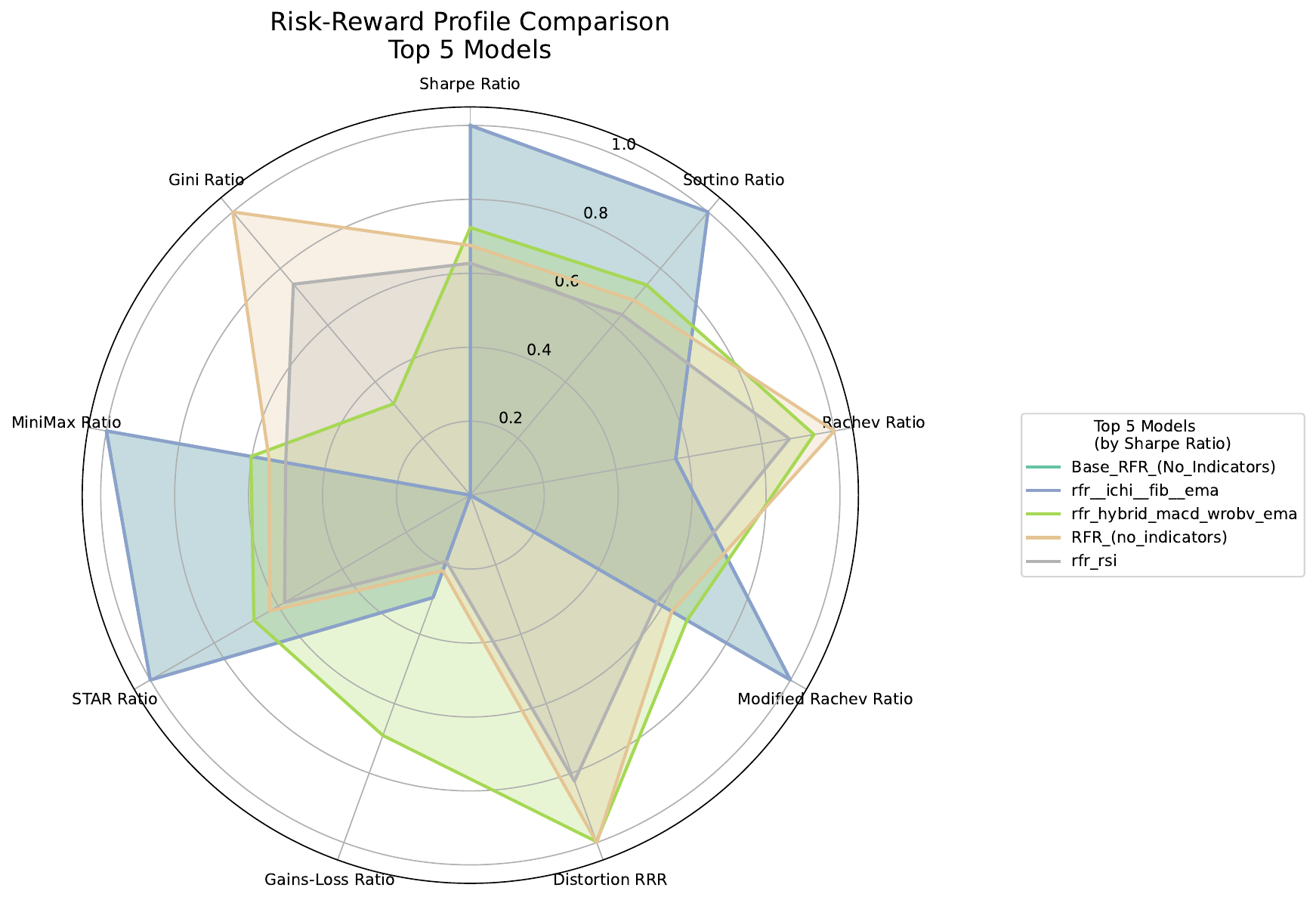}
\caption{Risk--reward profiles for the top five models, presenting the Sharpe, Sortino, and Rachev ratios.}
\label{fig:risk_reward_radar}
\end{figure}

\begin{figure}[htbp]
\centering
\includegraphics[width=\textwidth]{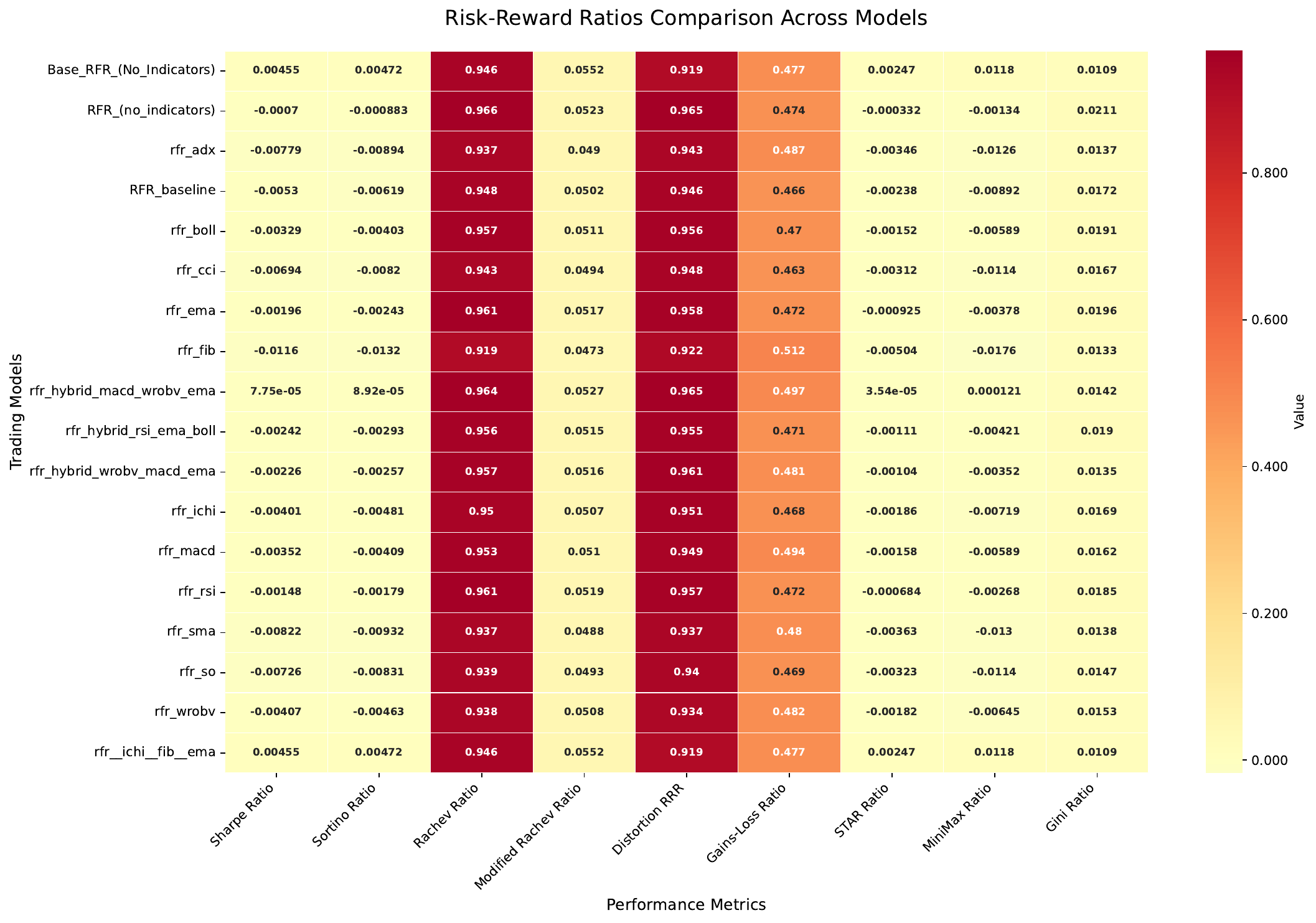}
\caption{Heatmap comparing Sharpe, Sortino, Rachev, and modified Rachev ratios for all models.}
\label{fig:risk_reward_heatmap}
\end{figure}

We assessed each strategy using risk metrics such as the Sharpe, Sortino, and Rachev ratios.
Figure~\ref{fig:risk_reward_radar} presents a radar chart comparing the top five models; Figure~\ref{fig:risk_reward_heatmap} presents a heatmap of their risk--reward profiles.

Despite the differing results  of the different strategies, none of the models yielded Sharpe ratios above 0.0046, a figure significantly below industry standards for viable trading strategies.
This suggests that technical indicators alone may not be sufficient for high-frequency trading, as the models struggle to achieve risk-adjusted returns that justify frequent trading.
Furthermore, the consistently negative Sortino ratios highlight that these models do not effectively protect against downside risk, reinforcing the argument that ML-based strategies in high-frequency trading environments face structural challenges.

Among the tested models, RFR\_RSI and RFR\_ICHIMOKU obtained slightly better Rachev ratios (0.919–0.961), suggesting that momentum-based indicators may offer a small advantage in risk--reward trade-offs.
However, these improvements were marginal and probably not statistically significant, indicating the need for further research with larger datasets and multi-asset testing.

A deeper analysis of the strategy’s performance reveals that most trading losses occurred during periods of heightened market volatility, suggesting that the models struggle to adapt dynamically to shifting volatility regimes.
These findings indicate that future research should explore adaptive models that adjust their feature weighting based on changing market conditions.
The inability to effectively navigate volatility shocks highlights a key limitation of static ML models in financial applications, reinforcing the need for more flexible approaches that can integrate real-time volatility estimation.

These results challenge the weak form of the Efficient Market Hypothesis (EMH), suggesting that technical indicators may contribute to risk-adjusted decision-making but fail to generate persistent excess returns.
This aligns with prior studies that found short-term inefficiencies in financial markets to be highly transient and difficult to exploit systematically.
Future work should examine whether alternative data sources, such as order book data, sentiment analysis, or macroeconomic signals, could enhance the predictive power.

\subsection{Feature Importance}\label{sec:feature_importance}

\subsubsection{Base Model}
\begin{figure}[htbp]
\centering
\includegraphics[width=0.8\textwidth]{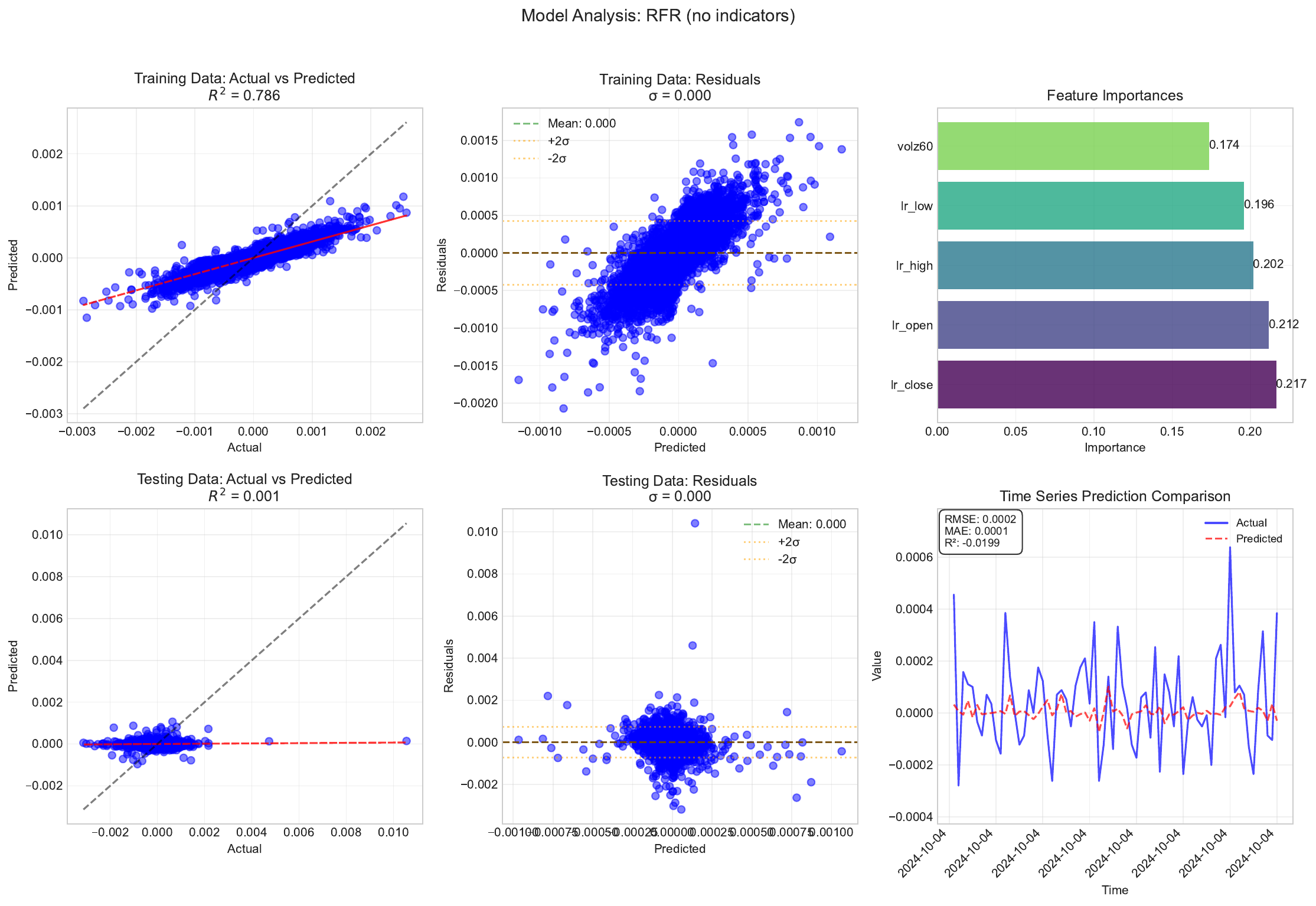}
\caption{Feature importance distribution for the base RFR model without technical indicators.}
\label{fig:base_importance}
\end{figure}

In the base RFR model (Figure~\ref{fig:base_importance}), the closing, opening, high, low, and volume (normalized as a $Z$-score) features accounted for over 90\% of the total importance.
This indicates that raw price and volume data captured most short-term market signals for minute-level trading, consistent with the literature suggesting that in high-frequency contexts, market noise overwhelms many of the usual indicators. 

\subsubsection{Technical Indicators}
\begin{figure}[htbp]
\centering
\includegraphics[width=0.8\textwidth]{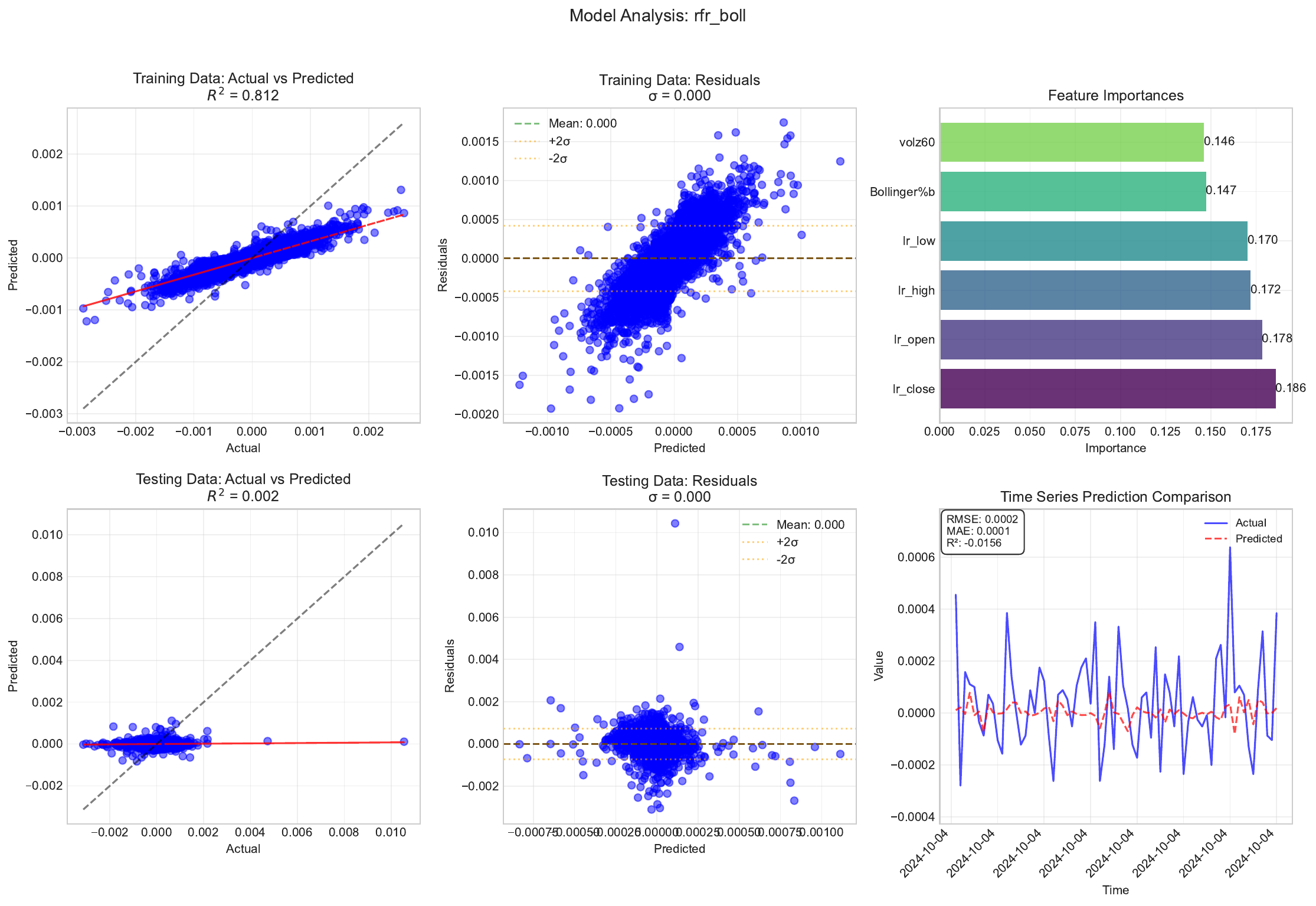}
\caption{Feature importance distribution for the RFR model including Bollinger Bands.}
\label{fig:boll_importance}
\end{figure}

\begin{figure}[htbp]
\centering
\includegraphics[width=0.8\textwidth]{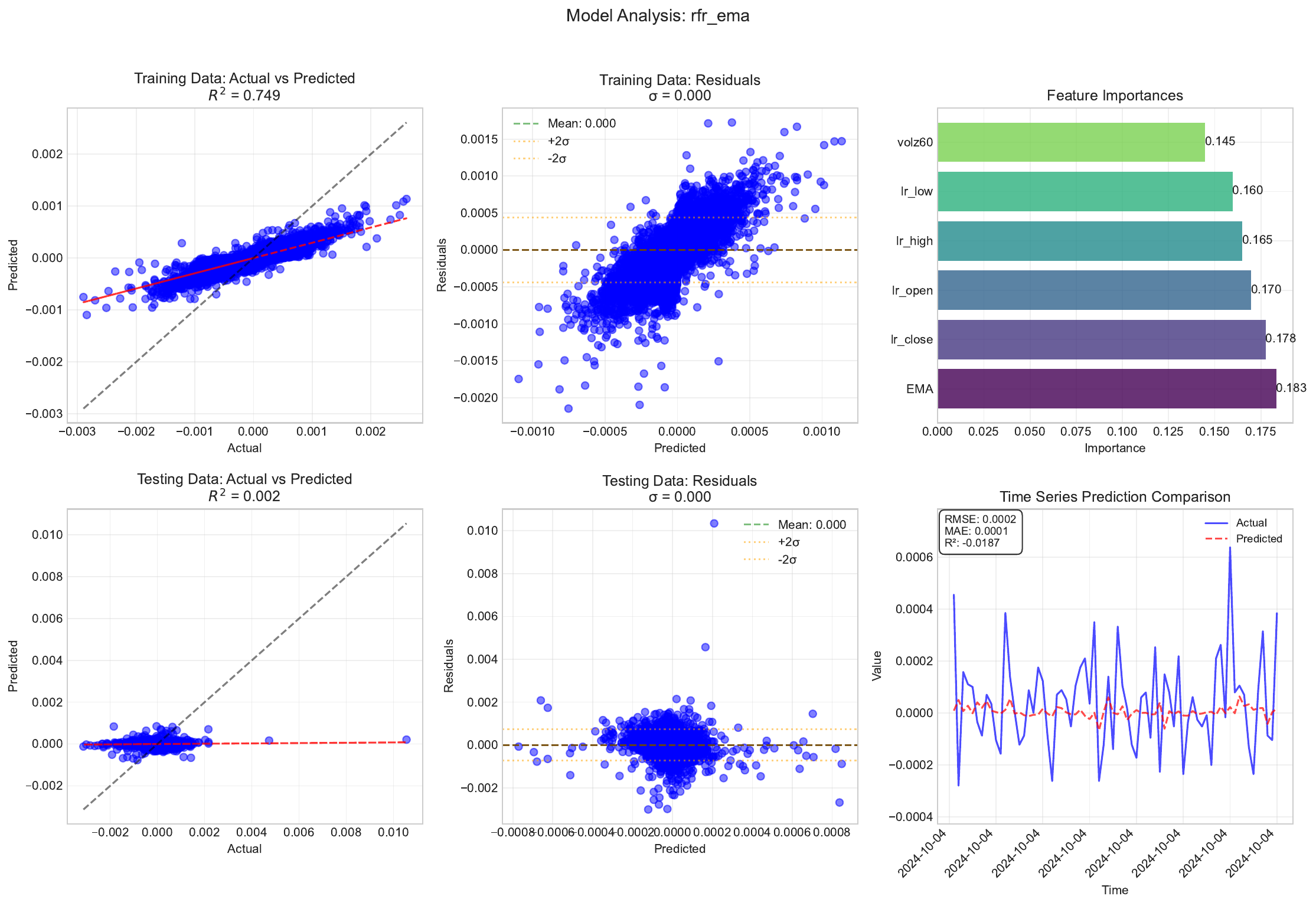}
\caption{Feature importance distribution for the RFR model including EMA.}
\label{fig:ema_importance}
\end{figure}

\begin{figure}[htbp]
\centering
\includegraphics[width=0.8\textwidth]{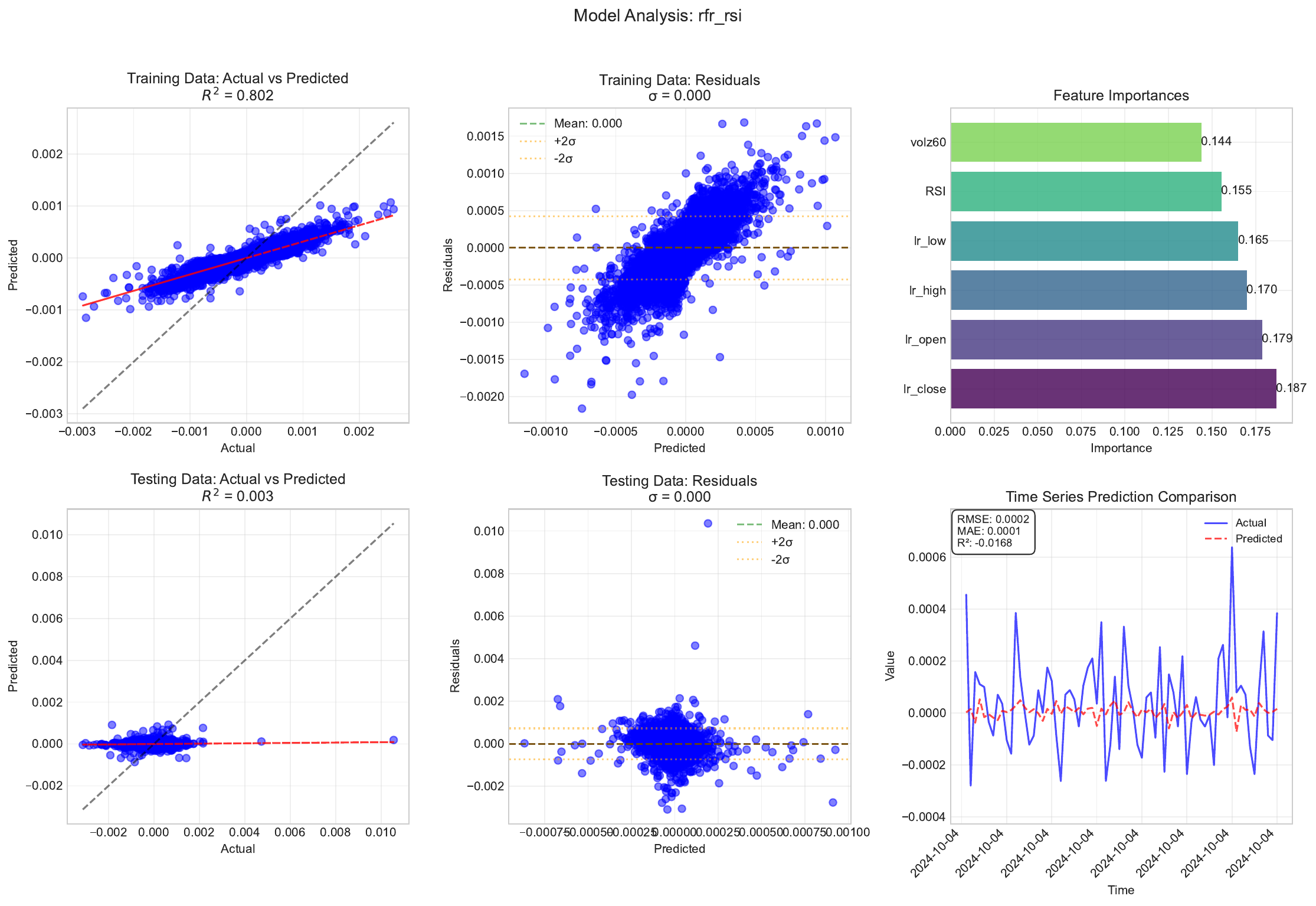}
\caption{Feature importance distribution for the RFR model including RSI.}
\label{fig:rsi_importance}
\end{figure}

Adding Bollinger Bands, EMA, or RSI (Figures~\ref{fig:boll_importance}--\ref{fig:rsi_importance}) shifted the distribution slightly, with these indicators contributing 14\%--18\% to the predictive decisions.
However, none of these changes substantially improved the out-of-sample accuracy or trading outcomes, implying that traditional indicators do not offer a stable advantage at minute-level frequency.

\subsection{Residual Analysis and Directional Accuracy}\label{sec:residual_and_accuracy}

\begin{figure}[htbp]
\centering
\includegraphics[width=\textwidth]{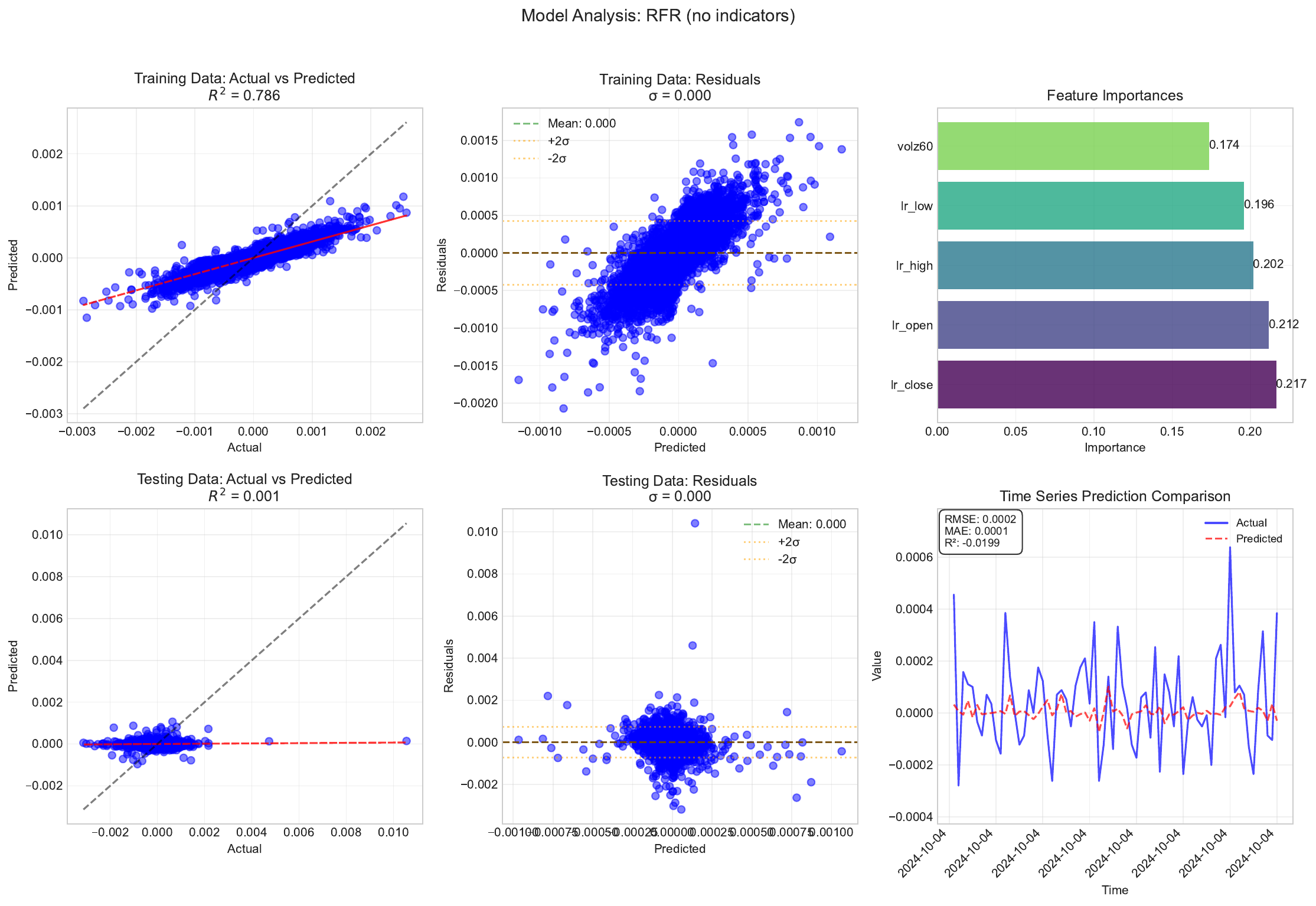}
\caption{Residual analysis for the base RFR model: actual vs.\ predicted returns and residual distributions during training and testing.}
\label{fig:residual_analysis}
\end{figure}

Residual plots for the base model (Figure~\ref{fig:residual_analysis}) showed no strong bias or autocorrelation, suggesting consistent performance within the sample.
However, the directional accuracy dropped notably from 80\%--87\% in training to 48\%--50\% in testing, again pointing to overfitting.
The correlation coefficients also declined from 0.86--0.92 (training) to 0.03--0.06 (testing).

\subsection{Comparative Analysis and Statistical Significance}\label{sec:comparative_analysis}

When comparing models that use standard technical indicators (e.g., RSI, EMA, and Bollinger Bands) with those relying only on features based on the raw prices, the former did not exhibit a clear advantage in out-of-sample prediction or final returns.
Although hybrid approaches combining multiple indicators slightly reduced the maximum drawdowns, they still failed to outperform simpler RFR models in absolute or risk-adjusted returns.

Statistical tests reinforced these findings: despite high in-sample \(R^2\), all models obtained negative out-of-sample \(R^2\).
Consistent RMSE and MAE values across variants of the model further suggest that adding technical indicators did not meaningfully reduce forecast errors.

In summary, these results highlight the challenges in exploiting minute-level data with standard technical indicators.
While the models fit historical data reasonably well, they struggled to generalize, indicating that high-frequency signals may be overshadowed by market noise and short-term volatility.
\section{Conclusion}

This study investigated the integration of technical indicators into Random Forest Regression (RFR) models for high-frequency stock price prediction, emphasizing both predictive accuracy and risk-adjusted performance.
Using minute-level SPY data, we systematically evaluated a range of technical indicators, including Bollinger Bands, Exponential Moving Averages (EMA), and Fibonacci retracements, to assess their contributions to the performance of the model under volatile market conditions.
The choice of SPY, a highly liquid and representative market proxy, ensures that our findings retain their significance for broader high-frequency trading applications.

Our results indicate that while technical indicators enhance certain risk-adjusted metrics, such as the Rachev and gains--loss ratios, their contribution to out-of-sample predictive accuracy remains limited.
A feature importance analysis consistently highlighted the dominance of primary price-based features (e.g., opening, closing, and high prices) over derived technical indicators.
Hybrid strategies incorporating multiple indicators demonstrated slight improvements in managing tail risks but failed to outperform the buy-and-hold benchmark in terms of returns.
These findings suggest that traditional technical indicators may have diminishing predictive value in modern high-frequency markets, where price discovery is driven primarily by raw price movements rather than widely recognized indicators.

Beyond predictive accuracy, this study advances the field by integrating advanced risk--reward measures to evaluate the practical viability of  trading strategies based on machine learning (ML).
While past research has focused predominantly on return maximization, our results emphasize the trade-offs between risk management and profitability.
The observed difficulties in generalization, where models exhibit strong in-sample performance but deteriorate significantly in out-of-sample testing, highlight the need for parsimonious modeling approaches that prioritize robustness over complexity.
This aligns with the existing literature on ML in financial markets, which finds overfitting to be a fundamental limitation in high-frequency trading applications.

From a theoretical standpoint, our findings provide insights into market efficiency and the feasibility of exploiting short-term price inefficiencies.
While the inability of our models to consistently generate excess returns aligns with the weak form of the Efficient Market Hypothesis (EMH), the ability of certain indicator-augmented strategies to maintain stable risk--reward profiles suggests that transient inefficiencies may persist under specific market conditions.
These results contribute to ongoing discussions on the microstructures of the market and the role of ML in financial decision-making.

Several challenges remain, including overfitting, the need for adaptive modeling techniques, and the computational costs associated with complex hybrid strategies.
Future research should explore dynamic, regime-aware models capable of adjusting to evolving market conditions while maintaining their computational efficiency.
Incorporating sources of alternative data, such as sentiment analysis and order book dynamics, could further enhance the predictive performance and provide deeper insights into price formation mechanisms.

From a practitioner’s perspective, this study highlights the importance of balancing interpretability, computational feasibility, and predictive power in the deployment of ML models for high-frequency trading.
While RFR-based strategies may not be optimal for maximizing absolute returns, their ability to manage tail risks and provide interpretable outputs means they can be valuable tools for risk-aware trading strategies.
Furthermore, technical indicators, such as Fibonacci retracement and the Ichimoku Cloud, despite their limited predictive power, may still have some practical utility due to their alignment with intuitive trading heuristics.

In conclusion, this study contributes to the growing body of literature on ML in financial markets by providing a nuanced assessment of the role of technical indicators in high-frequency trading.
While traditional indicators may have limited standalone predictive value, their integration within a structured risk-aware framework offers insights into market behavior and portfolio risk management.
Future research should focus on adaptive hybrid approaches that address the challenges to generalization, leverage sources of alternative data, and optimize computational efficiency, to enhance the practical applicability of ML in modern financial markets.

\section{Code Availability}

The implementation code for this study is available at \url{https://github.com/akashdeepo/ML_TI_RFR}.
The repository includes the core implementation files, \texttt{stockdata.py} for data processing and technical indicators, \texttt{pred\_rfr.py} for the Random Forest model, \texttt{simulate\_trading.py} for trading simulation, and \texttt{metrics.py} for performance evaluation.

The implementation uses the following Python libraries:
\begin{itemize}
    \item \texttt{scikit-learn}: Random Forest implementation with \texttt{RandomForestRegressor}
    \item \texttt{pandas} and \texttt{numpy}: Data manipulation and numerical computations
    \item \texttt{matplotlib} and \texttt{seaborn}: Visualization and plotting
    \item \texttt{logging}: Comprehensive logging for debugging and tracking
    \item Custom modules:
    \begin{itemize}
        \item Technical indicator computation
        \item Trading simulation with position sizing and turnover constraints
        \item Risk--reward ratio calculations including Rachev and Modified Rachev ratios
    \end{itemize}
\end{itemize}

The implementation emphasizes computational efficiency and real-time processing capabilities, with particular attention to high-frequency trading considerations.
The complete implementation requires access to minute-level SPY data through a Bloomberg Terminal subscription.
Users wishing to replicate this study should have appropriate Bloomberg Terminal access and the necessary subscriptions.
The code is provided as is under the MIT license, with the understanding that data acquisition and licensing compliance are the user's responsibility.

\section{Future Work}

While this study provides valuable insights into the role of technical indicators in high-frequency stock price prediction, several avenues remain open for further research.
A key limitation of this study is its focus on a single asset, the SPY.
Although the SPY was chosen for its high liquidity and broad market representation, future research should extend this analysis to multiple assets or multi-asset portfolios to evaluate the generalizability of the findings.
Expanding the study to diverse asset classes, such as commodities, fixed-income securities, and cryptocurrencies, would provide a deeper understanding of how technical indicators interact with varying market structures, liquidity conditions, and volatility regimes.

Another important direction is the integration of additional data sources to enhance the predictive performance and risk assessment.
Order book dynamics, sentiment analysis from financial news and social media, and alternative data sources such as macroeconomic indicators, could improve the feature selection and provide more context for trading decisions.
Investigating how these factors influence the performance of a model in high-frequency environments may yield more robust trading strategies.

Further, advances in deep learning architectures present an opportunity to capture complex sequential dependencies in high-frequency financial data.
Future studies should explore models such as Long Short-Term Memory (LSTM) networks and Transformer-based architectures, which have demonstrated strong performance in time-series forecasting tasks.
Additionally, comparisons with alternative ML techniques, such as gradient boosting methods or hybrid ensemble models, could provide insights into the optimal modeling approaches for different market conditions.

Finally, the challenges to practical implementation must be addressed to ensure the viability of ML-driven trading strategies in real-world applications.
Future research should explore the development of adaptive frameworks that dynamically adjust to evolving market regimes while incorporating real-world constraints, such as transaction costs, latency, and execution risks.
The integration of reinforcement learning techniques so as to optimize the execution  of trades and risk management strategies could further enhance the applicability of ML models in high-frequency trading.

By pursuing these research directions, future studies can contribute to the development of more resilient, interpretable, and efficient ML models for financial markets, ultimately bridging the gap between theoretical advances and practical deployment in trading environments.
\vspace{6pt}



\end{document}